\documentclass[floatfix,reprint,amsmath,amssymb,aps,pra]{revtex4-1}
\usepackage{times,mathptmx}
\usepackage{graphicx}
\usepackage[english]{babel}
\usepackage{epstopdf}
\usepackage{longtable}
\usepackage{xcolor}

\begin{document}

\title{Unambiguous scattering matrix for 
non-Hermitian systems}

\author{Andrey~Novitsky$^{1}$}
\email{novitsky@bsu.by}
\author{Dmitry~Lyakhov$^{2}$}
\author{Dominik~Michels$^{2}$}
\author{Alexander~A.~Pavlov$^{3}$}
\author{Alexander~S.~Shalin$^4$}
\author{Denis~V.~Novitsky$^{4,5}$}

\affiliation{ $^1$Department of Theoretical Physics and Astrophysics, Belarusian State University, Nezavisimosti Avenue 4, 220030 Minsk, Belarus \\ 
$^2$Visual Computing Center, King Abdullah University of Science and Technology, Thuwal 23955-6900, Kingdom of Saudi Arabia\\ 
$^3$ Institute of Nanotechnology of Microelectronics of the Russian Academy of Sciences, Leninsky Prospekt 32A, 119991 Moscow, Russia \\
$^4$ITMO University, Kronverksky Prospekt
49, 197101 St. Petersburg, Russia \\ $^5$B.~I.~Stepanov Institute of
Physics, National Academy of Sciences of Belarus, Nezavisimosti
Avenue 68, 220072 Minsk, Belarus}

\date{\today}

\begin{abstract}
$\mathcal{PT}$ symmetry is a unique platform for light manipulation and versatile use in unidirectional invisibility, lasing, sensing, etc. Broken and unbroken $\mathcal{PT}$-symmetric states in non-Hermitian open systems are described by scattering matrices. A multilayer structure, as a simplest example of the open system, has no certain definition of the scattering matrix, since the output ports can be permuted. The uncertainty in definition of the exceptional points bordering $\mathcal{PT}$-symmetric and $\mathcal{PT}$-symmetry-broken states poses an important problem, because the exceptional points are indispensable in applications as sensing and mode discrimination. Here we derive the proper scattering matrix from the unambiguous relation between the $\mathcal{PT}$-symmetric Hamiltonian and scattering matrix. We reveal that the exceptional points of the scattering matrix with permuted output ports are not related to the $\mathcal{PT}$ symmetry breaking. Nevertheless, they can be employed for finding a lasing onset as demonstrated in our time-domain calculations and scattering-matrix pole analysis. Our results are important for various applications of the non-Hermitian systems including encircling exceptional points, coherent perfect absorption, $\mathcal{PT}$-symmetric plasmonics, etc.
\end{abstract}

\maketitle

\section{Introduction} 

Scattering is a powerful tool for
investigating diverse phenomena in Nature. Due to the
interaction of a probe particle (electron, neutron, photon,
etc.) with a target, the former changes its characteristics that
can be measured by detectors. Independent of the type of the probe particle, the description of such an open system can be performed via the scattering matrix connecting input and output channels \cite{Newton}. 

One of the demanded applications of the scattering matrices is the parity-time ($\mathcal{PT}$)-symmetric systems, implying invariance under
simultaneous inversion of the spatial coordinates and time
\cite{Bender1998,Bender2007}. A $\mathcal{PT}$-symmetric system is an example of
a non-Hermitian system with the Hamiltonian $\hat H \neq \hat
H^\dag$, where the superscript $\dag$ stands for the Hermitian
conjugation. The $\mathcal{PT}$-symmetric non-Hermitian system can have a real-valued
spectrum of the operator $\hat H = \hat p^2/2m + \hat V$, if the
potential meets the condition $V({\rm r}) = V^\ast(-{\rm r})$.

Being introduced in the framework of quantum mechanics, the $\mathcal{PT}$
symmetry has attracted a lot of attention in photonics recently (see
the review articles
\cite{Zyablovsky2014,Feng2017,El-Ganainy2018,Ozdemir2019, Miri2019}). Using
the analogy between the Helmholtz and Schr\"odinger equations, one
can determine optical system's Hamiltonian with the real-valued
spectrum. Dielectric permittivity $\varepsilon$ plays the role of
the potential in photonics. That is why the $\mathcal{PT}$ symmetry requires
$\varepsilon({\rm r}) = \varepsilon^\ast(-{\rm r})$ in optics, i.e.
loss and gain should be balanced. Remarkably, such systems can be
fabricated in different designs: the couple of waveguides
\cite{Ruter2010}, the two-dimensional array of waveguides
\cite{Regensburger2012, Kremer2019}, the passive multilayer
\cite{Feng2014}, and the coupled meta-atom systems
\cite{Lawrence2014}. Along with an anisotropic transmission
\cite{Ge2012} and unidirectional invisibility
\cite{Lin2011,Feng2013}, intriguing physics arises in the vicinity
of exceptional points, where a phase transition between $\mathcal{PT}$- and
non-$\mathcal{PT}$-symmetric phases occurs \cite{Ozdemir2019,Miri2019}. The exceptional
points can be used as a platform for an optical omnipolarizer
\cite{Hassan2017} and sensor with unprecedented sensibility
\cite{Chen2017,Hodaei2017}. In the $\mathcal{PT}$-symmetry-broken state, a
system has exponentially decaying and amplifying modes, thus giving
an opportunity for lasing \cite{Feng2014-2,Hodaei2014,Gu2016} and
coherent perfect absorption \cite{Longhi2010,Chong2011,Wong2016}.

Scattering matrix can be written in terms of the Hamiltonian $\hat H$ as
$\hat S = \exp\left( - i \hat H (t-t_0)/\hbar \right)$, where $\hbar$ is the reduced Planck constant. It relates
the input wave at the moment $t_0$ with the output wave at the
moment $t$. For the real-valued spectrum $\lambda_n$ of the
$\mathcal{PT}$-symmetric Hamiltonian, we naturally obtain the unimodular
eigenvalues $s_n = \exp(-i \lambda_n (t-t_0)/\hbar)$ of the
scattering matrix: $|s_n| = 1$. Thus, the eigenvalues $s_n$
usually fully define, whether the state of a system is $\mathcal{PT}$-symmetric
or not. However, this may not be the case, if the scattering matrix
is introduced irrespective of the Hamiltonian. Then the input and
output channels can be permuted resulting in the scattering matrices
with different eigenvalues. Uncertainty in
the definition of the scattering matrix is inadmissible and can mislead during interpretation of results. Such situation arises for
one-dimensional (1D) multilayer structures shown in Fig. \ref{Fig1}. In this paper, we analyze the
definitions of the scattering matrix widely used in $\mathcal{PT}$-symmetry
studies, discuss their physical meaning and successfully solve the problem of
unique choice of the scattering matrix.

\section{Problem of the scattering matrix uniqueness} 

\begin{figure}[t!]
\centering \includegraphics[scale=1, clip=]{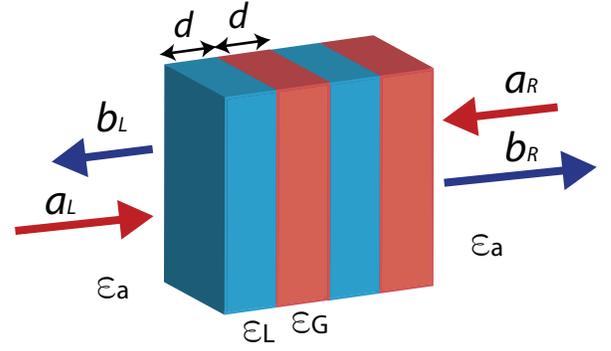}
\caption{\label{Fig1} $\mathcal{PT}$-symmetric multilayer system with
alternating loss and gain slabs. We employ parameters $d = 50$ nm
and $\varepsilon_a = 1$ through the paper. }
\end{figure}

Scattering
matrix of a 1D $\mathcal{PT}$-symmetric multilayer system schematically shown in
Fig. \ref{Fig1} is usually defined as
\begin{equation}
\hat S_1 = \left( \begin{array}{cc} r_L & t \\ t & r_R
\end{array} \right) \quad {\rm or} \quad \hat S_2 = \left(
\begin{array}{cc} t & r_R \\ r_L & t \end{array} \right),
\label{2defS}
\end{equation}
where $t$, $r_L$, and $r_R$ are the complex transmission,
reflection-to-the-left, and reflection-to-the-right coefficients,
respectively. The definitions differ in how we write the vector of
output fields: either $(b_L, b_R)^T = \hat S_1 (a_L, a_R)^T$ or
$(b_R, b_L)^T = \hat S_2 (a_L, a_R)^T$, where the superscript $T$
means the transposition, $a_{L,R}$ and $b_{L,R}$ are the input and
output fields, respectively (see Fig. \ref{Fig1}). It seems quite
not important, which matrix should be used, because they result in
the same output fields. However, this difference appears
dramatically important, when we are looking for the exceptional
points bordering the $\mathcal{PT}$-symmetric and $\mathcal{PT}$-symmetry-broken states.
Indeed, the eigenvalues $s_{1,2}$ of the
scattering matrix should be unimodular ($|s_{1,2}|=1$) in the $\mathcal{PT}$-symmetric state and inverse ($|s_1|=1/|s_2|$) in the $\mathcal{PT}$-symmetry-broken phase. The two definitions Eq. (\ref{2defS}) of the scattering matrix provide different
eigenvalues $s_{1,2}^{(1)} = (r_R+r_R)/2 \pm \sqrt{t^2 + (r_L -
r_R)^2/4}$ and $s_{1,2}^{(2)} = t \pm \sqrt{r_L r_R}$ for $\hat S_1$ and $\hat S_2$, respectively. Therefore,
the violation of $\mathcal{PT}$ symmetry appears under different conditions.
This situation is confusing and unacceptable.

The scattering matrix $\hat S_2$ is used in most articles on $\mathcal{PT}$
symmetry. Diverse phenomena have been described with its help
including unidirectional invisibility \cite{Lin2011}, optical
pulling \cite{Alaee2018}, and photonic spin Hall effect
\cite{Zhou2019}. 
Ambiguity of the definition of the scattering
matrix was discussed in several works \cite{Ge2012,Mostafazadeh}. In
particular, L. Ge \textit{et al.} in Ref. \cite{Ge2012} advocate for
$\hat S_1$ justifying this by the decisive comparison. In
particular, the exceptional points obtained using the scattering
matrix $\hat S_1$ are in good correspondence to the lasing onset
associated to breaking the loss-gain balance in $\mathcal{PT}$-symmetric
structures \cite{Ge2012}. This feature was used to predict the
properties of the systems acting simultaneously as a laser and
coherent perfect absorber \cite{Achilleos2017, Chong2011,Novitsky2019}, as well as
to study the lasinglike nonstationary behavior above the $\hat S_1$
exceptional point \cite{Novitsky2018}. Nevertheless, the matrix
$\hat S_1$ is rarely exploited in comparison to $\hat S_2$.

\section{Balance of the loss and gain} 

\begin{figure*}[t!]
\centering \includegraphics[scale=0.9, clip=]{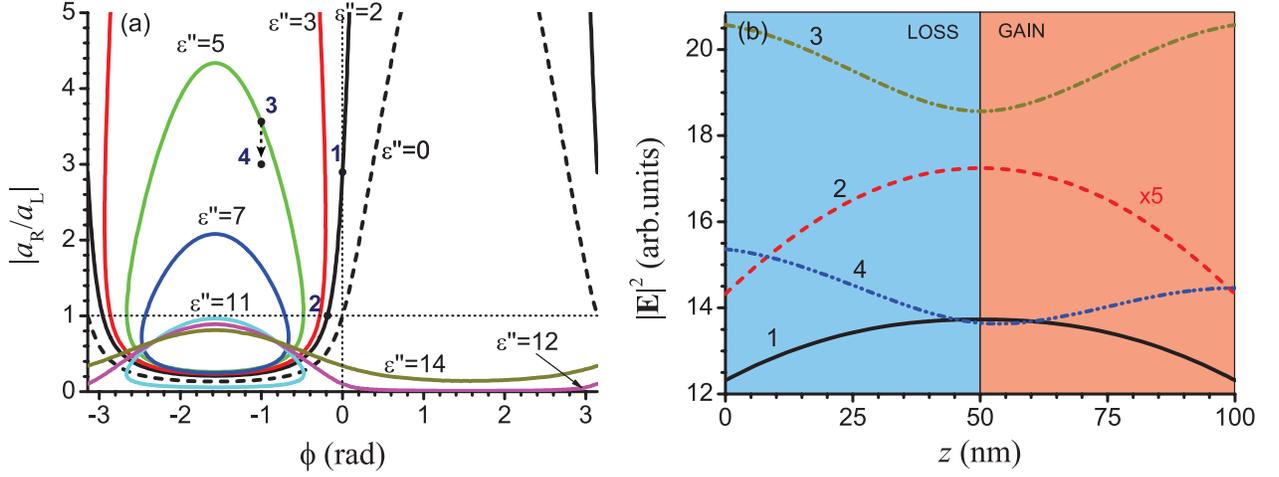}
\caption{\label{Fig2} (a) Curves of the loss and gain balance
(difference of the energy in loss and gain layers) for different
$\varepsilon''$ in a bilayer. (b) Energy density across the bilayer
for points 1 to 4 in (a). $|a_R|$ and $\phi$ are the absolute value
and argument of the complex amplitude $a_R$ at fixed $a_L = 1$ The
real part of permittivity is $\varepsilon' = 2$.}
\end{figure*}

$\mathcal{PT}$-symmetric phase is characterized by the balance of loss and gain. Therefore, first, we try to solve the dilemma ``$\hat S_1$ vs. $\hat S_2$'' using the assumption that only
a proper scattering matrix guarantees such a balance. The balance can be quantified as a zero
difference between the energies concentrated in the loss and gain
layers: $\Delta w = \left| \int_0^d |{\bf E}_L|^2 dz - \int_d^{2d}
|{\bf E}_G|^2 dz \right|$. The energy balance is a property of the system and, therefore, can be calculated irrespective of the scattering matrix. To this end, we specify one of the input amplitudes
$a_L = |a_L|$ and find the complex value $a_R$ to reach the balance. Since $a_R$ is characterized by the absolute value
$|a_R|$ and the argument $\phi$ (in fact, the phase difference with
$a_L$), we plot condition of the balance (at given $\varepsilon''$)
as curves in the coordinates $(|a_R/a_L|,\phi)$ in Fig.
\ref{Fig2}(a). For a scattering matrix, the amplitudes take certain values being its eigenvector $(a_L,a_R)^T$. These amplitudes correspond to a point on the curve of the energy balance. In fact, for the scattering matrix $\hat S_1$ the energy balance
is achieved at $|a_R| = |a_L|$, i.e. right and left input fields
differ only in phase [point 2 in Fig. \ref{Fig2}(a)]. The loss and
gain for the scattering matrix $\hat S_2$ are balanced, when the
left and right input fields are in phase, $\phi = 0$, and have
different magnitudes $|a_R| \neq |a_L|$ [point 1 in Fig. \ref{Fig2}(a)].
We emphasize, however, that the balance of loss and gain appears not
only at the points 1 and 2, but occupy a curve. Each point on the
curve have a symmetric energy distribution as demonstrated in Fig.
\ref{Fig2}(b) for points 1, 2, and 3 (the latter is calculated at $\varepsilon'' =
5$). Point 4 is displaced from the proper curve and, therefore, a
non-symmetric field distribution is seen in this case. Since the
scattering matrices are related as $\hat S_2 = \left(
\begin{array}{cc} 0 & 1 \\ 1 & 0 \end{array} \right) \hat S_1$, we
can assume that the loss and gain balance in point 3 might
correspond to another scattering matrix $\hat S_3 = B \hat S_1$,
where $B$ is an auxiliary matrix. Then there are infinitely many
scattering matrices that can be obtained from $\hat S_1$ using the
multiplication by this auxiliary matrix. So, the study of balance
conditions cannot solve the problem of the proper scattering matrix.


\section{Derivation of the scattering matrix} 

In quantum mechanics, a
system is $\mathcal{PT}$-symmetric, if its Hamiltonian $\hat H$ commutes with
the parity $\hat P$ and time $\hat T$ operators. Such a
non-Hermitian Hamiltonian can possess real eigenvalues $\lambda$,
which can be found from the stationary Schr\"odinger equation, $\hat
H \phi = \lambda \phi$. Time evolution of such a system follows from
the temporal Schr\"odinger equation $i \hbar
\partial \psi/\partial t = \hat H \psi$. If $\hat H$ is time
independent, one arrives at
\begin{equation}
\psi({\bf r}, t) = \exp\left( - \frac{i \hat H t}{\hbar} \right)
\psi({\bf r}, 0).
\end{equation}
This solution means that the initial state $\psi({\bf r}, 0)$ is
transformed into $\psi({\bf r},t)$ after some time $t$. The operator
connecting the wave functions before and after scattering is a
scattering matrix $\hat S = \exp\left( - i \hat H t/\hbar \right)$.
Commutation $[\hat S, \hat H]=0$ results in the same set of the
eigenfunctions for these operators: $\hat S \phi = \mu \phi$, where
$\mu = \exp(-i \lambda t/\hbar)$. For a system in the $\mathcal{PT}$-symmetric state, $\lambda$ is real-valued, hence, $\mu$ is unimodular, $|\mu| = 1$.
It is important that the scattering operator is unambiguously
determined by the Hamiltonian. We aim to find the scattering matrix for
multilayer systems in a similar manner.

Let us consider 1D propagation of a plane electromagnetic wave
through a $\mathcal{PT}$-symmetric multilayer system with dielectric
permittivity $\varepsilon(z) = \varepsilon^\ast(-z)$. The 1D
propagation implies dependence of the electric ${\bf E}(z,t)$ and magnetic ${\bf H}(z,t)$ fields on the single spatial
coordinate $z$ in the direction of the unit vector ${\bf q}$ normal to the slabs. Then the Maxwell equations can be formally written in the form of the temporal
Schr\"odinger equation
\begin{equation}
i\hbar \frac{\partial}{\partial t} \left( \begin{array}{c} {\bf
H}(z,t) \\ {\bf E}(z,t) \end{array} \right) = \hat H \left(
\begin{array}{c} {\bf H}(z,t) \\ {\bf E}(z,t) \end{array} \right),
\end{equation}
where the Hamiltonian is equal to
\begin{equation}
\hat H = i \hbar c {\bf q} \times \left( \begin{array}{cc} 0 & -1 \\
\varepsilon^{-1}(z) & 0 \end{array} \right) \frac{\partial}{\partial
z}.
\end{equation}

For the $\mathcal{PT}$-symmetric multilayer, the Hamiltonian is invariant under
the simultaneous space and time inversion. The scattering matrix
$\hat S = \exp\left( - i \hat H t/\hbar \right)$ connects the fields
before and after scattering as
\begin{equation}
\left( \begin{array}{c} {\bf H}(t) \\ {\bf E}(t) \end{array} \right)
= \hat S \left( \begin{array}{c} {\bf H}(0) \\ {\bf E}(0)
\end{array} \right). \label{HESHE}
\end{equation}
Without loss of generality further we assume a specific polarization. The fields before
scattering are the input fields $H_y(0) = H_y^{LR}(0) + H_y^{RL}(0)$
and $E_x(0) = E_x^{LR}(0) + E_x^{RL}(0)$: from the left to right
$H_y^{LR}(0)$, $E_x^{LR}(0) = \gamma_a H_y^{LR}(0)$ and from the
right to left $H_y^{RL}(0)$, $E_x^{RL}(0) = -\gamma_a H_y^{RL}(0)$,
where $\gamma_a = 1/\sqrt{\varepsilon_a}$ is the surface impedance.
Similarly, the output fields $H_y(t) = H_y^{LR}(t) + H_y^{RL}(t)$
and $E_x(t) = E_x^{LR}(t) + E_x^{RL}(t)$ after the scattering read
$H_y^{LR}(t)$, $E_x^{LR}(t) = \gamma_a H_y^{LR}(t)$ and
$H_y^{RL}(t)$, $E_x^{RL}(t) = -\gamma_a H_y^{RL}(t)$.

Introducing $a_L = H_y^{LR}(0)$, $a_R = H_y^{RL}(0)$, $b_L =
H_y^{RL}(t)$, and $b_R = H_y^{LR}(t)$, we get
\begin{equation}
\left( \begin{array}{c} b_R + b_L \\ \gamma_a (b_R - b_L)
\end{array} \right) = \hat S \left( \begin{array}{c} a_L + a_R \\
\gamma_a (a_L - a_R) \end{array} \right). \label{b+b=Sa+a}
\end{equation}
The scattering matrix can be written in terms of the reflection
$r_L$, $r_R$ and transmission $t$ coefficients. By substituting the
output fields $b_L = r_L a_L + t a_R$ and $b_R = r_R a_R + t a_L$
into Eq. (\ref{b+b=Sa+a}), we have
\begin{equation}
\hat S = \left( \begin{array}{cc} t + (r_R + r_L)/2 & -\gamma_a^{-1}(r_R-r_L)/2 \\
\gamma_a (r_R-r_L)/2 & t - (r_R + r_L)/2 \end{array} \right).
\label{S=t_r}
\end{equation}
This scattering matrix is unambiguously connected to the
$\mathcal{PT}$-symmetric Hamiltonian and, therefore, correctly describes
exceptional points and $\mathcal{PT}$-symmetry violation. The matrix (\ref{S=t_r})
can be presented as $\hat S = A \hat S_2 A^{-1}$, where $A =
-\frac{1}{2\gamma_a} \left( \begin{array}{cc} 1 & 1 \\
\gamma_a & -\gamma_a \end{array} \right)$ is the orthogonal rotation
matrix. That is why the eigenvalues of the scattering matrices $\hat
S$ and $\hat S_2$ coincide. Thus, the matrix $\hat S_2$ can be used
for correct prediction of the exceptional points, where the $\mathcal{PT}$
symmetry gets broken.

The derivation of the scattering matrix (\ref{S=t_r}) discussed
above can be supported by additional arguments. If the field is
incident from the left, the input field is characterized solely by
the amplitude $a_L$, while the output fields equal $b_L = r_L a_L$
and $b_R = t a_L$. Plugging them into Eq. (\ref{HESHE}) yields
\begin{equation}
\hat S \left( \begin{array}{c} 1 \\
\gamma_a \end{array} \right) = \left( \begin{array}{c} t + r_L
\\ \gamma_a (t - r_L) \end{array} \right). \label{Saltern}
\end{equation}
When the field comes in from the right-hand side, we have the
analogous equation, where $r_L$ and $\gamma_a$ are replaced with
$r_R$ and $-\gamma_a$, respectively:
\begin{equation}
\hat S \left( \begin{array}{c} 1 \\
-\gamma_a \end{array} \right) = \left( \begin{array}{c} t + r_R
\\ -\gamma_a (t - r_R) \end{array} \right). \label{Saltern1}
\end{equation}
Eqs. (\ref{Saltern}) and (\ref{Saltern1}) represent 4 equations for 4 unknown elements of the scattering matrix. Solution of these equations gives the same Eq. (\ref{S=t_r}).

\section{Physical meaning of the exceptional points}
\subsection{General remarks}
Using the generalized unitarity relation $r_L r_R = t^2(1 - |t|^{-2})$, the eigenvalues of the scattering matrices can be represented as $s_{1,2} = q (1 \pm \sqrt{1-|q|^{-2}})$, where $q = (r_L + r_R)/2$ for $\hat S_1$ and $q = t$ for $\hat S_2$ and $\hat S$. Then the $\mathcal{PT}$-symmetric state corresponds to $|q|^2<1$ resulting in $|s_{1,2}| = 1$. For $|q|^2>1$, we
arrive at the $\mathcal{PT}$-symmetry-broken state with $|s_1| = 1/|s_2|$.


\begin{figure}[t!]
\centering \includegraphics[scale=1, clip=]{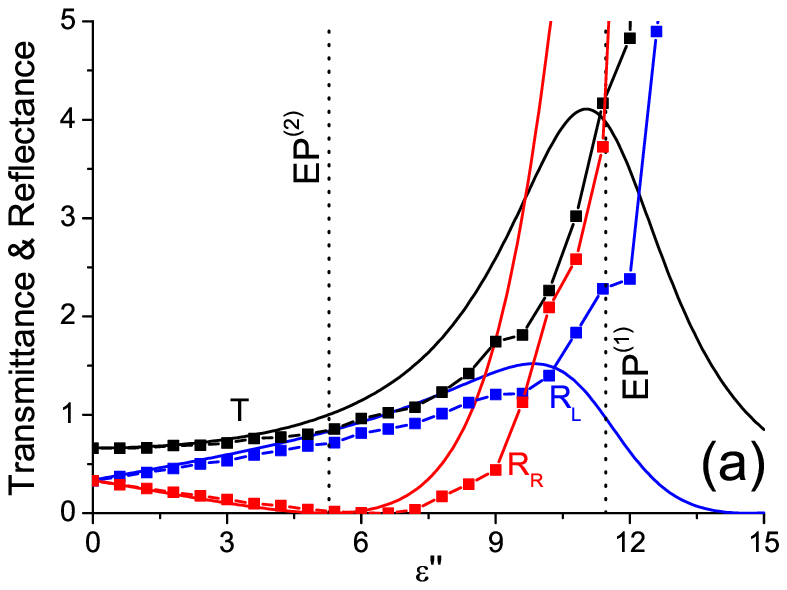}
\includegraphics[scale=1, clip=]{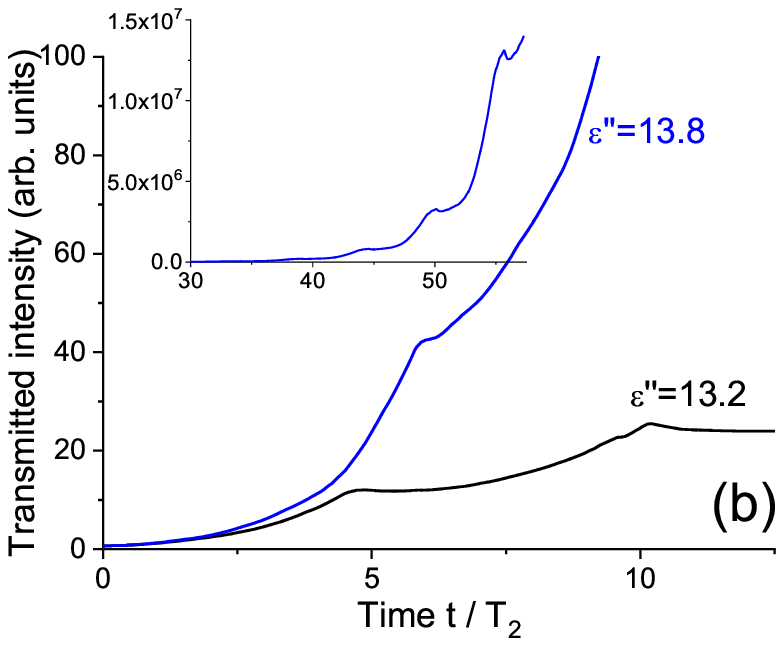}
\caption{\label{Fig3} (a) Transmittance $T$ and reflectances $R_L$ and $R_R$ of the loss-gain bilayer with $\varepsilon' = 4$ and $d=50$ nm as a function of $\varepsilon''$. The results of calculation using the transfer-matrix method and Maxwell-Bloch equations are shown by the solid lines and symbols, respectively. Vertical dotted lines mark positions of the exceptional points calculated using the scattering matrices $\hat S_1$ (EP$^{(1)}$) and $\hat S_2$ (EP$^{(2)}$). (b) Temporal profiles
of transmitted intensity for $\varepsilon''$ just below and above the lasing threshold near the EP$^{(1)}$. The inset shows the curve for $\varepsilon''=13.8$ on the longer scale.}
\end{figure}


$\mathcal{PT}$-symmetric phase for $\hat S_2$ corresponds to $|t|^2 < 1$, while exceptional points follow from $|t| = 1$. The exceptional points of the matrix $\hat S_1$ are defined by the
condition $|r_L + r_R| = 2$. Remarkably, it can be met for the great
values of reflectances $|r_L|^2$ and $|r_R|^2$, if the difference of arguments $|{\rm Arg}(r_{L}) - {\rm Arg}(r_{R})|$ is around $\pi$. So, the $\mathcal{PT}$ symmetry breaking via
$\hat S_1$ may occur near poles of the reflection and transmission coefficients and can be used as a predictor of lasing as
discussed in Refs. \cite{Ge2012,Chong2011,Novitsky2018}. 

\subsection{Time-domain simulations}

Further we shed light on the role of the exceptional points of $\hat S_1$ using time-domain calculations of the
wave propagation through the bilayer with parameters
$\varepsilon' = 4$ and $d=50$ nm. Treating the loss and gain materials as two-level media, we numerically solve the Maxwell-Bloch equations \cite{Novitsky2011} for the complex amplitude of
microscopic (atomic) polarization $\rho$, difference between
populations of the ground and excited states $w$, and electric field amplitude $A$:
\begin{eqnarray}
\frac{d\rho}{d\tau}&=& i l \Omega w + i \rho \delta - \gamma_2 \rho, \label{dPdtau} \\
\frac{dw}{d\tau}&=&2 i (l^* \Omega^* \rho - \rho^* l \Omega) -
\gamma_1 (w-w_{eq}),
\label{dNdtau} \\
\frac{\partial^2 \Omega}{\partial \xi^2}&-& n_d^2 \frac{\partial^2
\Omega}{\partial \tau^2}+2 i \frac{\partial \Omega}{\partial \xi}+2
i n_d^2 \frac{\partial \Omega}{\partial
\tau} + (n_d^2-1) \Omega \nonumber \\
&&=3 \alpha l \left(\frac{\partial^2 \rho}{\partial \tau^2}-2 i
\frac{\partial \rho}{\partial \tau}-\rho\right), \label{Maxdl}
\end{eqnarray}
where $\tau=\omega t$ and $\xi=kz$ are respectively the
dimensionless time and distance, $\Omega=(\mu/\hbar \omega) A$ is
the normalized Rabi frequency, $\omega$ is the light circular frequency, $k = \omega/c$ is the
wavenumber in vacuum, $c$ is the speed of light, $\hbar$ is the
reduced Planck constant, and $\mu$ is the dipole moment of the
quantum transition. Dimensionless parameter $\alpha= \omega_L /
\omega = 4 \pi \mu^2 C/3 \hbar \omega$ is the light-matter coupling
strength, where $\omega_L$ is the Lorentz frequency and $C$ is the
concentration of two-level atoms. $\delta=(\omega_0-\omega)/\omega$
is the detuning of the light frequency $\omega$ from the frequency
$\omega_0$ of the atomic resonance. Normalized relaxation rates
of population $\gamma_1=1/(\omega T_1)$ and polarization
$\gamma_2=1/(\omega T_2)$ are expressed by means of the longitudinal
$T_1$ and transverse $T_2$ relaxation times. Influence of the
polarization of the background dielectric having real-valued
refractive index $n_d=\sqrt{\varepsilon'}$ on the embedded active
particles is taken into account by the local-field enhancement
factor $l=(n_d^2+2)/3$ \cite{Crenshaw2008,Bloembergen}.

Equilibrium population difference $w_{eq}$ plays the role of a
pumping parameter: it changes from $w_{eq}=1$ (no pump, purely lossy
medium) via $w_{eq}=0$ (no loss or gain, saturated medium) to
$w_{eq}=-1$ (maximal pump, fully inverted medium). In the
steady-state approximation, when a saturation can be neglected
($\Omega \ll \Omega_{sat}=\sqrt{\gamma_1 (\gamma_2^2+\delta^2)/4l^2
\gamma_2}$), one can obtain
the expression for the effective permittivity of the two-level
medium at the exact resonance $\delta=0$ \cite{Novitsky2017}: $\varepsilon_{eff} = \varepsilon' + i
\varepsilon'' \approx n_d^2+3 i l^2 \omega_L T_2 w_{eq}$. We
assume that the $\mathcal{PT}$-symmetric bilayer is composed of a lossy ($\varepsilon_{eff+}$) and gain ($\varepsilon_{eff-}$) layers, where \cite{Novitsky2018}
\begin{eqnarray}
\varepsilon_{eff\pm} \approx n_d^2 \pm 3 i l^2 \omega_L T_2
|w_{eq}|. \label{epsPT}
\end{eqnarray}
The system is obviously $\mathcal{PT}$-symmetric, because the
necessary condition $\varepsilon(z) = \varepsilon^\ast(-z)$ is satisfied.

Equations (\ref{dPdtau})--(\ref{Maxdl}) are solved numerically using
the finite-difference approach developed in Ref.
\cite{Novitsky2009}. An initial value of the population difference
is $w(t=0)=w_{eq}$. In our calculations, we assume $\delta=0$ (exact
resonance), $n_d=2$, $\omega_L=10^{11}$ rad$\cdot$s$^{-1}$, $T_1=1$ ns, and
$T_2=50$ ps. These parameters correspond to the semiconductor quantum dots as the active particles \cite{Palik, Diels}. The loss and gain layers have the same thickness $d=50$ nm. The incident monochromatic light has the wavelength $\lambda=1$ $\mu$m and the intensity much lower than the saturation intensity. These parameters might
correspond, e.g., to the quantum dots as two-level atoms, but
the other systems with properly
chosen parameters can be also considered. The side-pumping scheme similar to that realized
by Wong \textit{et al} \cite{Wong2016} can be employed for excitation of the materials.

In Fig. \ref{Fig3}(a), we compare the transmittance and
reflectance calculated using the time-domain (TD) and transfer-matrix (TM)
techniques. The positions of the exceptional points EP$^{(1)}$ and EP$^{(2)}$ for the corresponding scattering matrices $\hat S_1$ and $\hat S_2$ divide the whole space of $\varepsilon''$ into three regions. There is a good agreement between the TD and TM methods at low $\varepsilon''$ (first region), including the position of EP$^{(2)}$ corresponding to the condition of a unidirectional reflectionlessness \cite{Ge2012}.
Between the exceptional points EP$^{(1)}$ and EP$^{(2)}$ (second region), there is a divergence between 
the results given by the TD and TM methods, in accordance
with the previous observations \cite{Novitsky2018}. Both reflectances and transmittance are larger for the TM approach. There are several reasons for this discrepancy. First, the linearized expression for permittivity (\ref{epsPT}) used in TM calculations gives less correct results for larger gain and loss, since it does not take into account the nonlinear effect of saturation. As a result, the TM method neglecting nonlinearity gives somewhat overestimated transmission and reflection in comparison to the TD simulations. Second, the spatial discretization used in the TD simulations can influence the calculations for large gain, since the amplification of radiation per every spatial step becomes more significant. Nevertheless, the general trends of transmission and reflection change are consistent for both calculation methods.

Above the EP$^{(1)}$ (third region), the situation dramatically changes, because the system loses
stability and goes into the lasinglike mode described within the TD technique. The development of lasing can be
more vividly observed in the temporal dynamics illustrated in
Fig. \ref{Fig3}(b): a comparatively small increase in
$\varepsilon''$ results in the transfer from the regime of a stationary state
establishment to exponentially increasing intensity. Limited by the saturation, the
exponential increase results in the formation of a powerful pulse as analyzed in Ref.
\cite{Novitsky2018}. Here we show only proof-of-concept results and the values of $\varepsilon''$
necessary for observing these effects can be made much lower for the greater number of layers and thicker slabs.

Thus, from the dynamical considerations, we deduce three regimes of the light interaction: (i) the $\mathcal{PT}$-symmetric state with balanced loss and gain below the EP$^{(2)}$, (ii) the $\mathcal{PT}$-symmetry-broken state with imbalanced loss and gain, but
nevertheless stationary levels of reflection and transmission
(between EP$^{(1)}$ and EP$^{(2)}$), and (iii) the $\mathcal{PT}$-symmetry-broken state with nonstationary (lasinglike) mode of radiation above EP$^{(1)}$. Wave propagation in a $\mathcal{PT}$-symmetric system emulates propagation in lossless medium with $T<1$, when the loss and gain compensate each other. This qualitatively explains that $T>1$ requires violation of the $\mathcal{PT}$ symmetry in agreement with the $\hat S_2$ behavior. To the contrast, lasing phenomenon is not connected with the $\mathcal{PT}$ symmetry breaking. This latter regime needs to be discussed in more detail.

\subsection{Lasing onset and pole analysis}


\begin{figure}[t!]
\includegraphics[scale=1, clip=]{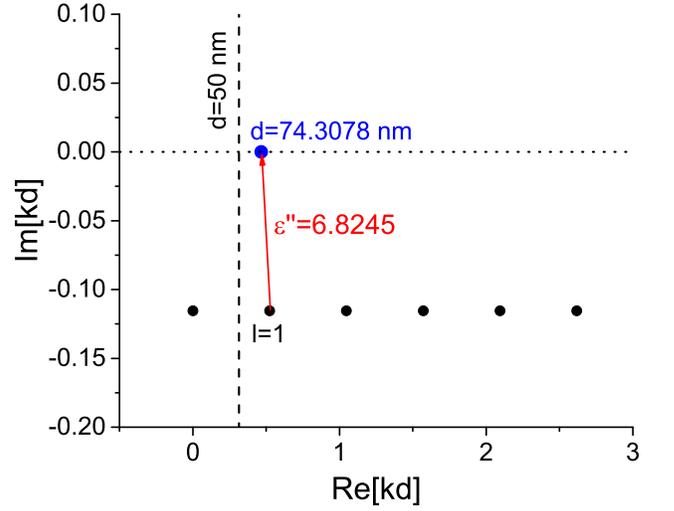}
\caption{\label{Fig4} The positions of the poles in the complex-frequency plane. The cases of the lossless bilayer (black dots) and the bilayer with $\varepsilon'' = 6.8245$ (blue dot) are shown. }
\end{figure}


Lasing is usually associated with poles of the scattering matrix arising at the real frequencies, where both reflection and transmission coefficients tend to infinity. In the case of a bilayer with refractive indices $n_L=\sqrt{\varepsilon_L}$ and $n_G=\sqrt{\varepsilon_G}$, positions of the poles in the complex frequency plane can be found semianalytically from equation
\begin{eqnarray}
&& \left( \frac{n_L}{n_G} + \frac{n_G}{n_L} \right) \tan{kdn_L} \tan{kdn_G} \nonumber \\ 
&+& i \left[ \left( n_L + \frac{1}{n_L} \right) \tan{kdn_L} + \left( n_G + \frac{1}{n_G} \right) \tan{kdn_G} \right] \nonumber \\
&-& 2 = 0, \label{poles}
\end{eqnarray}
where $kd=2 \pi d / \lambda$ is the normalized frequency. Solving Eq. (\ref{poles}), one obtains the positions of the poles on the complex-frequency plane $({\rm Re} [kd], {\rm Im} [kd])$. The lossless structure (real $n_L$, $n_G$) has a series of poles located in the lower half of the complex-frequency plane, i.e. at ${\rm Im} [kd] < 0$ (see the black dots in Fig. \ref{Fig4}). Introducing loss and gain, one can shift these singularities towards the real axis. When the pole reaches the real axis, so that ${\rm Im} [kd] = 0$, the system starts lasing. The corresponding value of $\varepsilon''$ is usually treated as a lasing threshold. In Fig. \ref{Fig4}, we show a trajectory of the first non-zero pole ($l=1$) that appears at the real axis at $\varepsilon'' = 6.8245$. For the parameters considered above ($\varepsilon' = 2$, $\lambda=1$ $\mu$m), we obtain the necessary thickness of the layers as $d=74.3078$ nm. This value is larger than $d=50$ nm used in our time-domain simulations and marked with vertical dashed line in Fig. \ref{Fig4}.


\begin{figure}[t!]
\includegraphics[scale=1, clip=]{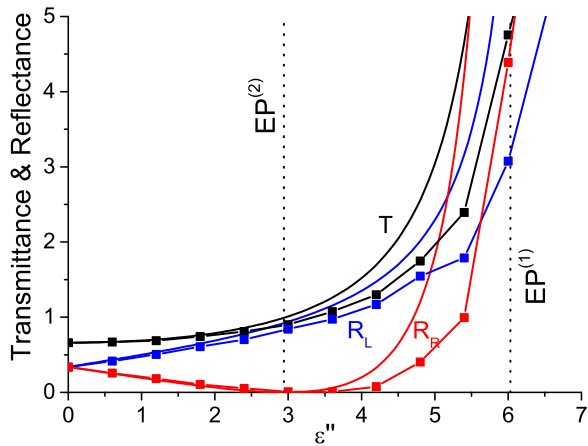}
\caption{\label{Fig5} The same as in Fig. \ref{Fig3}(a), but for $d=74.308$ nm (near the pole shown in Fig. \ref{Fig4}).}
\end{figure}


Thus, the conditions of the pole appearance are not met in Fig. \ref{Fig3}. Nevertheless, the lasinglike response is still feasible in this non-pole regime, though it requires a much stronger gain $\varepsilon'' > 13$ [see Fig. \ref{Fig3}(b)] that is almost twice as large compared with the lasing at the pole. To make the difference even more clear, in Fig. \ref{Fig5} we plot the same quantities as in Fig. \ref{Fig3}(a), but now at the pole. One can see that in contrast to the non-pole case, the transmittance and reflectance for the parameters corresponding to the pole show almost unbounded growth as we get closer to the EP$^{(1)}$. This is definitely associated with the lasing onset. 

To explain the behavior of waves in vicinity of the pole, we will make several remarks. First, it should be noted that we deal with an external wave interacting with the gain-containing system. This implies that the frequency of the radiation is externally prescribed and amplification and instability analysis can be performed only at this \textit{single} frequency. On the contrary, the ``true'' lasing develops from the radiation fluctuations at \textit{all} frequencies, while the lasing frequency is governed by the parameters of the structure (cavity) and the gain medium. Since the pole provides the lowest instability threshold, the lasing frequency usually corresponds to the pole conditions. Second, we stress that the availability of the poles is sufficient, but not necessary for the lasing onset in the case of external irradiation considered. This is absolutely reasonable, since the poles are the geometric (zero-dimensional) points, which can be reached only with the perfect tunability of the system parameters (in our case, the precise values of the frequency and layer thickness). The instability outside the singular points is achieved in non-optimal conditions (for larger gain), so that the reflectance and transmittance of the structure are not infinite, but reach certain large enough values.

Returning to the meaning of the exceptional points, we note that the exceptional points of $\hat S_1$ allow us to approximately predict the lasing threshold, in accordance with the instability of the light propagation shown in Fig. \ref{Fig3}(b). Since the transmittance in this case should be greater than $1$, the EP$^{(1)}$ appears at the larger $\varepsilon''$ than the EP$^{(2)}$. This means that the EP$^{(1)}$ can be used as a lasing predictor, even when the conditions of a pole appearance are not fulfilled.

\section{Conclusion} 

We have solved an important problem regarding the correct form of the scattering matrix of a photonic non-Hermitian system and, thus, determined conditions of the $\mathcal{PT}$ symmetry breaking. In particular, we have derived the scattering matrix of the 1D multilayer heterostructure unequivocally related to the $\mathcal{PT}$-symmetric Hamiltonian. It should be used for the correct prediction of the exceptional points (when transmittance $T=1$) and $\mathcal{PT}$-symmetry breaking (when $T>1$), solving the dilemma of the proper scattering matrix. Exceptional points of the scattering matrix with permuted output ports (when the sum of reflection coefficients $|r_L + r_R| = 2$) are of great importance, too, since they can roughly predict the lasinglike regime. The lasing has nothing to do with the symmetry breaking, but it is related to the development of instability in the dynamic system. The described theory is not limited by the multilayer structures, but can be adapted to any non-Hermitian open system. The discussed issues might be applied for analysis of the sensing and lasing designs possessing fascinating sensibility and modal structure.

\section*{Acknowledgements}

The work was supported by the Belarusian
Republican Foundation for Fundamental Research (Project No.
F18R-021), the Russian Foundation for Basic Research (Projects No.
18-02-00414, No. 18-52-00005 and No. 18-32-00160). Numerical simulations of light interaction with
resonant media were supported by the Russian Science Foundation
(Project No. 18-72-10127).


\begin{thebibliography}{0}

\bibitem{Newton} R.G. Newton, Scattering Theory of Waves and
Particles (Springer-Verlag Berlin Heidelberg, 1982).

\bibitem{Bender1998} C.~M.~Bender and S.~Boettcher,
Real spectra in non-Hermitian Hamiltonians having $\mathcal{PT}$
symmetry, {Phys. Rev. Lett.} \textbf{80}, 5243 (1998).

\bibitem{Bender2007} C.~M.~Bender,
Making sense of non-Hermitian Hamiltonians, {Rep. Prog. Phys.}
\textbf{70}, 947 (2007).

\bibitem{Zyablovsky2014} A.~A.~Zyablovsky, A.~P.~Vinogradov, A.~A.~Pukhov, A.~V.~Dorofeenko, and A.~A.~Lisyansky,
$\mathcal{PT}$ symmetry in optics, {Phys. Usp.} \textbf{57}, 1063
(2014).

\bibitem{Feng2017} L.~Feng, R.~El-Ganainy, and L.~Ge,
Non-Hermitian photonics based on parity-time symmetry, {Nat.
Photon.} \textbf{11}, 752 (2017).

\bibitem{El-Ganainy2018} R.~El-Ganainy, K.~G.~Makris, M.~Khajavikhan, Z.~H.~Musslimani, S.~Rotter, and D.~N.~Christodoulides,
Non-Hermitian physics and $\mathcal{PT}$ symmetry, {Nat. Phys.}
\textbf{13}, 11 (2018).

\bibitem{Ozdemir2019} S.K. \"Ozdemir, S. Rotter, F. Nori, and L. Yang,
Parity-time symmetry and exceptional points in photonics, Nat.
Mater. \textbf{18}, 783 (2019).

\bibitem{Miri2019} M.-A. Miri and A. Alu, Exceptional points in optics and photonics, Science \textbf{363}, eaar7709 (2019).

\bibitem{Ruter2010} C.~E.~R\"uter, K.~G.~Makris, R.~El-Ganainy,
D.~N.~Christodoulides, M.~Segev, and D.~Kip, Observation of
parity-time symmetry in optics, {Nat. Phys.} \textbf{6}, 192 (2010).

\bibitem{Regensburger2012} A. Regensburger, C. Bersch, M.-A. Miri,
G. Onishchukov, D.N. Christodoulides, and U. Peschel, Parity-time
synthetic photonic lattices, Nature, \textbf{488}, 167 (2012).

\bibitem{Kremer2019} M. Kremer, T. Biesenthal, L.J. Maczewsky, M.
Heinrich, R. Thomale, and A. Szameit, Demonstration of a
two-dimensional $\mathcal{PT}$-symmetric crystal, Nat. Commun. \textbf{10}, 435
(2019).

\bibitem{Feng2014} L. Feng, X. Zhu, S. Yang, H. Zhu, P. Zhang, X.
Yin, Y. Wang, and X. Zhang, Demonstration of a large-scale optical
exceptional point structure, Opt. Express \textbf{22}, 1760 (2014).

\bibitem{Lawrence2014} M.~Lawrence, N.~Xu, X.~Zhang, L.~Cong, J.~Han, W.~Zhang, and S.~Zhang,
Manifestation of $\mathcal{PT}$ symmetry breaking in polarization
space with Terahertz metasurfaces, {Phys. Rev. Lett.} \textbf{113},
093901 (2014).

\bibitem{Ge2012} L.~Ge, Y.~D.~Chong, and A.~D.~Stone,
Conservation relations and anisotropic transmission resonances in
one-dimensional $\mathcal{PT}$-symmetric photonic heterostructures,
{Phys. Rev. A} \textbf{85}, 023802 (2012).

\bibitem{Lin2011} Z.~Lin, H.~Ramezani, T.~Eichelkraut, T.~Kottos, H.~Cao, and D.~N.~Christodoulides,
Unidirectional invisibility induced by $\mathcal{PT}$-symmetric
periodic structures, {Phys. Rev. Lett.} \textbf{106}, 213901 (2011).

\bibitem{Feng2013} L. Feng, Y.-L. Xu, W. S. Fegadolli, M.-H. Lu, J. E. B. Oliveira, V.
R. Almeida, Y.-F. Chen, and A. Scherer, Experimental demonstration
of a unidirectional reflectionless paritytime metamaterial at
optical frequencies, \textit{Nat. Mater.} \textbf{12}, 108 (2013).

\bibitem{Hassan2017} A.~U.~Hassan, B.~Zhen, M.~Soljacic, M.~Khajavikhan, and D.~N.~Christodoulides,
Dynamically encircling exceptional points: Exact evolution and
polarization state conversion, {Phys. Rev. Lett.} \textbf{118},
093002 (2017).

\bibitem{Chen2017} W.~Chen, S.~K.~\"{O}zdemir, G.~Zhao, J.~Wiersig, and L.~Yang,
Exceptional points enhance sensing in an optical microcavity,
{Nature (London)} \textbf{548}, 192 (2017).

\bibitem{Hodaei2017} H.~Hodaei, A.~U.~Hassan, S.~Wittek, H.~Garcia-Gracia,
R.~El-Ganainy, D.~N.~Christodoulides, and M.~Khajavikhan, Enhanced
sensitivity at higher-order exceptional points, {Nature (London)}
\textbf{548}, 187 (2017).

\bibitem{Feng2014-2} L.~Feng, Z.~J.~Wong, R.-M.~Ma, Y.~Wang, and X.~Zhang,
Single-mode laser by parity-time symmetry breaking, {Science}
\textbf{346}, 972 (2014).

\bibitem{Hodaei2014} H.~Hodaei, M.-A.~Miri, M.~Heinrich, D.~N.~Christodoulides, and M.~Khajavikhan,
Parity-time-symmetric microring lasers, {Science} \textbf{346}, 975
(2014).

\bibitem{Gu2016} Z.~Gu, N.~Zhang, Q.~Lyu, M.~Li, S.~Xiao, and Q.~Song,
Experimental demonstration of $\mathcal{PT}$-symmetric stripe lasers, {Laser
Photon. Rev.} \textbf{10}, 588 (2016).

\bibitem{Longhi2010} S.~Longhi,
$\mathcal{PT}$-symmetric laser absorber, {Phys. Rev. A} \textbf{82},
031801(R) (2010).

\bibitem{Chong2011} Y.~D.~Chong, L.~Ge, and A.~D.~Stone,
Symmetry Breaking and Laser-Absorber Modes in Optical Scattering
Systems, {Phys. Rev. Lett.} \textbf{106}, 093902 (2011).

\bibitem{Wong2016} Z.~J.~Wong, Y.-L.~Xu, J.~Kim, K.~O'Brien, Y.~Wang, L.~Feng, and X.~Zhang,
Lasing and anti-lasing in a single cavity, {Nat. Photon.}
\textbf{10}, 796 (2016).

\bibitem{Alaee2018} R. Alaee, J. Christensen, and M. Kadic,
Optical Pulling and Pushing Forces in Bilayer $\mathcal{PT}$ -Symmetric
Structures, Phys. Rev. Appl. \textbf{9}, 014007 (2018).

\bibitem{Zhou2019} X. Zhou, X. Lin, Z. Xiao, T. Low, A. Alù, B. Zhang, and
H. Sun, Controlling photonic spin Hall effect via exceptional
points, Phys. Rev. B \textbf{100}, 115429 (2019).

\bibitem{Mostafazadeh} A. Mostafazadeh, Scattering Theory and
$\mathcal{PT}$-Symmetry, in \textit{Parity-time Symmetry and Its Applications}, ed. D. Christodoulides and J. Yang (Springer, 2018), pp. 75-121.

\bibitem{Achilleos2017} V. Achilleos, Y. Au\'regan, and V. Pagneux, Scattering by Finite Periodic $\mathcal{PT}$
-Symmetric Structures, Phys. Rev. Lett. \textbf{119}, 243904 (2017).

\bibitem{Novitsky2019} D.V. Novitsky, CPA-laser effect and exceptional points in $\mathcal{PT}$-symmetric multilayer structures, J. Opt.
\textbf{21}, 085101 (2019).

\bibitem{Novitsky2018} D.V. Novitsky, A. Karabchevsky, A.V. Lavrinenko, A.S. Shalin,
and A.V. Novitsky, $\mathcal{PT}$ symmetry breaking in multilayers with resonant
loss and gain locks light propagation direction, Phys. Rev. B
\textbf{98}, 125102 (2018).

\bibitem{Novitsky2011} D.~V.~Novitsky,
Femtosecond pulses in a dense two-level medium: Spectral
transformations, transient processes, and collisional dynamics,
{Phys. Rev. A} \textbf{84}, 013817 (2011).

\bibitem{Crenshaw2008} M.~E.~Crenshaw,
Comparison of quantum and classical local-field effects on two-level
atoms in a dielectric, {Phys. Rev. A} \textbf{78}, 053827 (2008).

\bibitem{Bloembergen} N.~Bloembergen, \textit{Nonlinear Optics} (Benjamin, New York, 1965).

\bibitem{Novitsky2017} D.~V.~Novitsky, V.~R.~Tuz, S.~L.~Prosvirnin, A.~V.~Lavrinenko, and A.~V.~Novitsky,
Transmission enhancement in loss-gain multilayers by resonant
suppression of reflection, {Phys. Rev. B} \textbf{96}, 235129
(2017).

\bibitem{Novitsky2009} D.~V.~Novitsky,
Compression of an intensive light pulse in photonic-band-gap
structures with a dense resonant medium, {Phys. Rev. A} \textbf{79},
023828 (2009).

\bibitem{Palik} E.~D.~Palik (ed.), \textit{Handbook of Optical Constants of Solids} (Academic Press, San Diego, 1998).

\bibitem{Diels} J.-C.~Diels and W.~Rudolph, \textit{Ultrashort Laser Pulse Phenomena} (Academic Press, San Diego, 2nd edn, 2006).

\bibitem{Wong2016} Z.~J.~Wong, Y.-L.~Xu, J.~Kim, K.~O'Brien, Y.~Wang, L.~Feng, and X.~Zhang,
Lasing and anti-lasing in a single cavity, {Nat. Photon.}
\textbf{10}, 796 (2016).


\end{thebibliography}
\end{document}